\documentclass[11pt]{article}
\pdfoutput=1
\usepackage[T1]{fontenc}
\usepackage[latin1]{inputenc}
\usepackage[nosort]{cite}
\usepackage{color}
\usepackage[bulletsep]{collref}
\usepackage{graphicx}
\usepackage{bbm}
\usepackage{amsmath}
\usepackage{amssymb}
\usepackage{subfig}
\usepackage{multirow}
\usepackage{tikz}
\usepackage{hyperref}

\usepackage{ytableau}

\makeatletter \@addtoreset{equation}{section} \makeatother

\makeatletter
\let\old@startsection=\@startsection
\let\oldl@section=\l@section
\renewcommand{\@startsection}[6]{\old@startsection{#1}{#2}{#3}{#4}{#5}{#6\mathversion{bold}}}
\renewcommand{\l@section}[2]{\oldl@section{\mathversion{bold}#1}{#2}}
\makeatother

\makeatletter
\let\old@makecaption=\@makecaption
\def\@makecaption{\small\old@makecaption}
\makeatother

\let\oldPhi=\Phi
\let\oldPsi=\Psi
\let\oldGamma=\Gamma
\let\oldDelta=\Delta
\let\oldSigma=\Sigma
\let\oldTheta=\Theta
\let\oldPi=\Pi
\let\oldUpsilon=\Upsilon
\renewcommand{\Phi}{\mathnormal{\oldPhi}}
\renewcommand{\Psi}{\mathnormal{\oldPsi}}
\renewcommand{\Gamma}{\mathnormal{\oldGamma}}
\renewcommand{\Sigma}{\mathnormal{\oldSigma}}
\renewcommand{\Delta}{\mathnormal{\oldDelta}}
\renewcommand{\Theta}{\mathnormal{\oldTheta}}
\renewcommand{\Pi}{\mathnormal{\oldPi}}
\renewcommand{\Upsilon}{\mathnormal{\oldUpsilon}}

\newcommand{\tr}{\mathop{\mathrm{tr}}}

\renewcommand{\Im}{\mathop{\mathrm{Im}}}

\newcommand{\Sphere}{\mathrm{S}}  
\newcommand{\AdS}{\mathrm{AdS}}

\ifx\genfrac\sdflkaj

\else

\fi

\newcommand{\p}{\partial}

\def\[{\begin{equation}}
\def\]{\end{equation}}

\makeatletter
\def\mr@ignsp#1 {\ifx\:#1\@empty\else #1\expandafter\mr@ignsp\fi}%
\newcommand{\multiref}[1]{\begingroup
\xdef\mr@no@sparg{\expandafter\mr@ignsp#1 \: }%
\def\mr@comma{}%
\@for\mr@refs:=\mr@no@sparg\do{\mr@comma\def\mr@comma{,}\ref{\mr@refs}}%
\endgroup}
\makeatother

\newcommand{\hypref}[2]{\ifx\href\asklfhas #2\else\href{#1}{#2}\fi}

\renewcommand{\eqref}[1]{(\multiref{#1})}

\ifx\href\asklfhas\newcommand{\href}[2]{#2}\fi

\newcommand{\be}{\begin{eqnarray}}
\newcommand{\ee}{\end{eqnarray}}

\DeclareMathOperator{\arcsinh}{arcsinh}

\newcommand{\sstar}{SYM${}^*$ }

\begin{document}

\thispagestyle{empty}
\begin{flushright}\footnotesize
\texttt{NORDITA-2015-XXX}\\
\texttt{UUITP-XX/15}\\
\end{flushright}
\vspace{1cm}

\begin{center}%
{\Large\textbf{\mathversion{bold}%
Holographic Wilson Loops in Symmetric Representations\\ in $\mathcal{N}=2^*$ Super-Yang-Mills Theory
}\par}

\vspace{1.5cm}

\textrm{Xinyi Chen-Lin${}^{1,2}$, Amit Dekel${}^{1}$ and Konstantin Zarembo${}^{1,2}$} \vspace{8mm} \\
\textit{%
${}^1$Nordita, KTH Royal Institute of Technology and Stockholm University, \\
Roslagstullsbacken 23, SE-106 91 Stockholm, Sweden\\
${}^2$Department of Physics and Astronomy, Uppsala University\\
SE-751 08 Uppsala, Sweden
} \\

\texttt{\\ xinyic@nordita.org, amit.dekel@nordita.org, zarembo@nordita.org}

\par\vspace{14mm}

\textbf{Abstract} \vspace{5mm}

\begin{minipage}{14cm}
We construct the D3-brane solution in the holographic dual of
the $\mathcal{N}=2^*$ theory that describes Wilson lines
in symmetric representations of the gauge group.
The results perfectly agree with the direct field-theory
predictions based on  localization.
\end{minipage}

\end{center}

\newpage

\tableofcontents

\bigskip
\noindent\hrulefill
\bigskip

\section{Introduction}\label{sec:intro}

The $\mathcal{N}=2^*$ super-Yang-Mills theory (SYM${}^*$), a mass deformation of $\mathcal{N}=4$ super-Yang-Mills preserving maximal amount of supersymmetry, is special in many respects. The gauge coupling in this theory does not run and is an adjustable parameter. The holographic dual of \sstar  is known explicitly as the Pilch-Warner (PW) solution of type IIB supergravity \cite{Pilch:2000ue}.  The curvature of this solution is controlled by the inverse of the 't~Hooft couping $\lambda =g_{\rm YM}^2N$, and the supergravity approximation is neatly justified when $\lambda $ is large. On the other hand, a number of non-perturbative results in
\sstar are available, in particular via localization of the path integral on $S^4$ \cite{Pestun:2007rz}, making possible directly comparison of field-theory results with holography in this non-conformal set-up.

Expectation values of large Wilson loops in SYM${}^*$, calculated via localization, were found to agree with the area law in the Pilch-Warner geometry at strong coupling \cite{Buchel:2013id}. The agreement was also established for the free energy on $S^4$ \cite{Bobev:2013cja}. Here we concentrate on Wilson loops in higher representations of $SU(N)$, whose rank $k$ scales with $N$ in the large-$N$ limit such that $k/N$ stays finite. Their expectation values were calculated in \cite{Chen-Lin:2015dfa} with the help of the systematic strong-coupling expansion of the localization partition function \cite{Chen:2014vka,Zarembo:2014ooa}. The strong-coupling regime of planar \sstar shows an interestingly irregular pattern, with infinitely many phase transitions accumulating towards $\lambda =\infty $ \cite{Russo:2013qaa,Russo:2013kea}, and Wilson loops in higher representations were suggested as probes sensitive to the non-trivial phase structure \cite{Chen-Lin:2015dfa}. The phase transitions however affect the expectation values of the Wilson loops at higher orders in $1/\sqrt{\lambda }$  \cite{Chen-Lin:2015dfa}. In this paper we study the leading-order, to begin with, which should not be affected by the strong-coupling irregularities.

The holographic dual of a Wilson loop in the rank-$k$ representation is a D-brane with $k$ units of electric flux on its world-volume. The Wilson loop/D-brane duality has been extensively studied in the AdS/CFT context \cite{Drukker:2005kx,Hartnoll:2006hr,Yamaguchi:2006tq,Gomis:2006sb,Hartnoll:2006is,RodriguezGomez:2006zz,Gomis:2006im,Yamaguchi:2007ps,Fiol:2014vqa,Faraggi:2011ge,Faraggi:2014tna,Buchbinder:2014nia}, allowing for a fairly detailed comparison of localization results with holography in the maximally superconformal case. Our goal is to generalize these results to the non-conformal setting of \sstar by
constructing D-brane embeddings in the Pilch-Warner geometry. In this paper we concentrate on D3-branes, which are dual to Wilson loops in symmetric rank-$k$ representations \cite{Gomis:2006im,Yamaguchi:2007ps}. Probe branes in the Pilch-Warner background have been studied in the past \cite{Buchel:2000cn,Evans:2000ct}, but with different boundary conditions that correspond to Higgsing the gauge group of the theory, rather than inserting a Wilson loop operator in the path integral.

\section{Review of the matrix model results}\label{sec:MM}

The Wilson loop in the $\mathcal{N}=2^*$ theory is defined as
\begin{equation}
 W_{\mathcal{R}}(C)=\left\langle \mathop{\mathrm{tr}}\nolimits_{\mathcal{R}}{\rm P}\exp\left[
 \int_{C}^{}ds\,\left(iA_\mu \dot{x}^\mu +\Phi \left|\dot{x}\right|\right)
 \right]\right\rangle,
\end{equation}
where $\mathcal{R}$ is an arbitrary representation of $SU(N)$ and $\Phi $ is one of the two adjoint scalar fields in the vector multiplet. In principle,  an arbitrary linear combination of the two adjoint scalars can appear in the Wilson loop, and the coupling can even change with the position on the contour, but localization can only compute Wilson loops with the constant coupling indicated above.

Localization computes the path integral of \sstar (or any other theory with $\mathcal{N}=2$ supersymmetry) compactified on the four-sphere by reducing it to a zero-dimensional matrix model \cite{Pestun:2007rz}. The Wilson loop expectation values and the free energy on $S^4$ have been computed by solving this model at large-$N$ and at strong coupling \cite{Buchel:2013id,Chen:2014vka,Zarembo:2014ooa} and has been successfully matched with the geometric predictions of (classical) string theory in the PW background \cite{Buchel:2013id,Bobev:2013cja}. Here we concentrate on Wilson loops in an arbitrary symmetric representation of $SU(N)$:
\begin{equation}
 \mathcal{R}_k=
 \overbrace{
 \begin{ytableau}
 ~ & ~ & ~ & \none & \none[\dots] & \none  &  \\
 \end{ytableau}}^{k}
\end{equation}

If $k\ll N$, large-$N$ factorization prescribes that
\begin{equation}\label{large-N-fac}
 W_{\mathcal{R}_k}(C)=\frac{1}{k!}\,\left[W(C)\right]^k,
\end{equation}
up to $1/N$ corrections.  This is no longer true for $k\sim N$, when the Wilson loop expectation value acquires a non-trivial dependence on $k/N$.  The natural scaling variable at large $\lambda $ is actually
\begin{equation}\label{kappa-def}
 \kappa =\frac{\sqrt{\lambda }\,k}{4N}\,.
\end{equation}
The limit we will concentrate upon is
\begin{equation}\label{scalinglaw-sugra}
 N\rightarrow \infty ,\qquad
 k\rightarrow \infty , \qquad
 \lambda \gg 1,\qquad \kappa {\rm \,-\,fixed.}
\end{equation}
The holographic dual of a Wilson loop in this regime is a classical D3-branes with $k$ units of electric flux on its world-volume.

The scaling limit that arises in the strong-coupling solution of the localization matrix model is slightly different. First of all, the path integral localizes only if the Wilson loop runs along the big circle of $S^4$, and strictly speaking this is the only contour amenable to exact calculation on the field theory side. But expectation values of sufficiently large Wilson loops should be universal, and largely independent of the detailed shape of the contour. In a massive $\mathcal{N}=2$ theory we expect the perimeter law, and indeed the logarithm of the circular Wilson loop in the fundamental representation scales linearly with the radius of the sphere. The coefficient in front of the perimeter term can therefore be read off  from the result for the circular loop on $S^4$. Following this logic, the localization prediction was successfully compared to the holographic calculation for the a long fragment of the straight line in $\mathbbm{R}^4$  \cite{Buchel:2013id}.

Wilson loops in higher representations have been computed from localization  in \cite{Chen-Lin:2015dfa}. The matrix model generates three disparate energy scales at strong coupling, namely $\sqrt{\lambda }M\gg M\gg 1/R$, where $R$ is the radius of the four-sphere and $M$ is the mass that appears in the Lagrangian of SYM${}^*$. Depending on the size of the representation, several scaling regimes are possible. The regime that resembles (\ref{scalinglaw-sugra}) most, and in fact includes it as a particular case, is sensitive to the largest of the mass scales above, $\sqrt{\lambda }M$. The localization path integral then simplifies a lot, essentially reducing to the Gaussian matrix model. The results for the Wilson loops in an arbitrary symmetric representation can then be directly transplanted from $\mathcal{N}=4$ SYM \cite{Hartnoll:2006is}, where the matrix model is Gaussian \cite{Erickson:2000af,Drukker:2000rr} from the very beginning \cite{Chen-Lin:2015dfa}:
\begin{eqnarray}\label{logW}
 &&\ln W_{\mathcal{R}_k}\simeq \frac{NM^2L^2}{2\pi ^2}\,f\left(\frac{\pi \sqrt{\lambda }\,k}{2MLN}\right)
\nonumber \\
&&f(x)=x \sqrt{1+x ^2}+\arcsinh x.
\end{eqnarray}
 The natural scaling regime on the gauge theory side is therefore:
\begin{equation}\label{scalinglaw-SYM}
 N\rightarrow \infty ,\qquad
 k\rightarrow \infty , \qquad
 \lambda \gg 1,\qquad \frac{\kappa}{ML} {\rm \,-\,fixed.}
\end{equation}

To match the two regimes, we need to take $\kappa \ll ML$ in (\ref{logW}). Then,
\begin{equation}\label{eq: MatrixModelResult}
 \ln W_{\mathcal{R}_k}\simeq \frac{\sqrt{\lambda }\,kML}{2\pi }\,,
\end{equation}
which should be valid in the regime (\ref{scalinglaw-sugra}). This simple prediction is a bit surprising, it just follows from the factorization formula (\ref{large-N-fac}) and indicates that
the D-brane with $k$ units of electric flux behaves exactly as a stack of $k$ non-interacting strings.

\section{D3-brane in the Pilch-Warner background}

The Pilch-Warner background is a deformation of $AdS_5\times S^5$, to which it reduces close to the boundary (in the UV). We review the supergravity solution in detail in the appendix~\ref{sec:PWB}.
 The D3-brane configuration dual to the $k$-symmetric Wilson loop sits at one point, i.e. $\theta=\pi/2 $ and $\phi=0$, on the deformed $S^5$.
At this point, the general Pilch-Warner metric (\ref{eq:PWmetric}) reduces to a deformed $AdS_5$ metric, which, in the string frame, is parametrized as:
\begin{align}\label{eq: deformedAdS5}
 ds^2 = \dfrac{A(c) M^2}{c^2-1}(dx^2 - d\rho^2 -\rho^2 d\Omega_2^2)
      -	\dfrac{1}{A(c)(c^2-1)^2}dc^2,
\end{align}
and
\begin{align}
A  = c+(c^2 - 1)\frac{1}{2}\ln\left(\frac{c-1}{c+1}\right).
\end{align}
Note that we use $c$ as coordinate instead of  $r$ originally used in \cite{Pilch:2003jg}. Near the boundary the coordinate $c$ is related to the standard holographic coordinate $z$ of $AdS_5$ in the Poincar\'e patch as
\begin{equation}
 c=1+\frac{z^2M^2}{2}\,.
\end{equation}
 We also stated explicitly the mass scale $M$ in the metric in order to compare with the result from the matrix model.

The world-volume of a D3-brane dual to a Wilson loop of contour $C$ has a boundary of the $S^1 \times S^2$ topology.
The non-contractible cycle of the boundary on the D-brane is mapped to the contour $C$ on the boundary of the holographic space-time.
The D3-brane embedding, near the boundary, has the $AdS_2 \times S^2$ geometry, where $S^2$ has asymptotically constant invariant volume.
Since the metric blows up near the boundary, the coordinate volume of $S^2$ shrinks to zero, and all the points of $S^1 \times S^2$  are actually mapped to the same one-dimensional contour $C$.

As the matrix model result is expected to be universal for any large contour, it is sufficient to study $C$ being an infinite line, in $x$ coordinate. 
The embedding consistent with the symmetry of the problem is parametrized by the identical map of $x$ and the two angles on $S^2$.
Then, we can choose either to have a non-trivial profile for $c = c(\rho)$ or $\rho = \rho(c)$.
We will choose the latter one, and hence, $c$ is also mapped identically to the world-volume.
The  metric induced on the D3-brane is
\begin{align}\label{eq: inducedMetric}
 ds_{\text{ind.}}^2 = \dfrac{A M^2}{c^2-1}\left(dx^2  - \rho^2 d\Omega_2^2 \right)
		      -\left(\dfrac{1}{A \left(c^2-1\right)^2} + \dfrac{A M^2 \rho'^2}{c^2-1} \right) dc^2,
\end{align}
where $\rho' = d\rho/dc$.

The goal is to find $\rho(c)$ and then, compute the D3-brane action in Euclidean signature, which is related to the dual Wilson loop at strong coupling, namely,
\[
 \ln W_{\mathcal{R}_k}= - S_{D3}.
\]

\subsection{D3-brane action}
The D3-brane action in the Euclidean signature is:
\begin{align}\label{eq:D3brane}
S = T_{D_3} \int_{\mathcal{M}} d\sigma^4 e^{-\Phi}
	      \sqrt{
		\det\limits_{ij} \left( g_{i j} + B_{i j} + \frac{1}{T_{F_1}} F_{i j} \right)
	      }
      - T_{D_3} \int_{\mathcal{M}} P[C_{(4)}]
      - i k \int_{\Sigma} F,
\end{align}
where $\mathcal{M}$ is the D3-brane world-volume,
$g_{i j}$ is the induced metric,
$B_{i j}$ is the pullback of the $B_{MN}$ field which vanishes at $\theta = \pi/2 $,
$F_{i j}$ is the internal gauge field on the D-brane world-volume,
and $\Sigma$ is a disk whose boundary is a non-contractible cycle on the boundary of the D-brane.
In our conventions, the metric is dimensionless and has unit radius near the boundary, therefore, the string tension is also dimensionless: $T_{F_1} = \sqrt{\lambda}/(2 \pi) $.
The D3-brane tension in these units is $T_{D_3} = N/(2\pi^2)$.

Note that, apart from the Dirac-Born-Infeld (DBI) action and the Wess-Zumino (WZ) term,
we added a Lagrange multiplier term that contains the amount of string charge $k$ that the D$p$-brane carries.
This latter arises from the coupling to the $B$ field, which, at linearized order,
\begin{align}
\delta S_{\text{DBI}} =  T_{F_1} \int d^{p+1}\sigma \Pi^{i j} \,  B_{i j}, \quad \Pi^{i j}=\dfrac{\delta S_{\text{DBI}}}{\delta F_{i j}},
\end{align}
must be compared with how the string is coupled to the $B$ field (the imaginary $i$ is due to the Euclidean signature):
\begin{align}
\delta S_{\text{str}} = T_{F_1} \int d^{2}\sigma \, \dfrac{i}{2} \epsilon^{i j} \, B_{i j}.
\end{align}
The D$p$-brane is locally $\Sigma \times S^{p-1}$, where $\Sigma$ is the string world-sheet.
The D-brane carries the correct string charge provided that the electric field has components only along $\Sigma$.
Upon averaging over the sphere $ S^{p-1}$, we obtain
\begin{align}
 \int_{S^{p-1}}d^{p-1}\sigma \Pi^{i j} = \dfrac{i \, k}{2} \epsilon^{i j},
\end{align}
which is achieved by adding the Lagrange multiplier to the D-brane action.

Back to the D3-brane action, the non-trivial component of the internal gauge field is the electric field $F_{x c}(c)$.
It is convenient to rescale $\rho\rightarrow \rho/M$, and introduce $ f= \frac{2\pi}{\sqrt{\lambda}M}F_{x c}$.
After integrating over $x$ and the 2-sphere, the D-brane action becomes,
\begin{align}\label{eq:D3action}
S = \frac{2 N M L}{\pi} \int dc
\left[
  \frac{c A \rho^2}{c^2 - 1}
  \sqrt{\frac{A^2 \rho'^2}{(c^2-1)^2}
    	+ \frac{1}{(c^2 - 1)^3}
	+ f^2}
  - \frac{c A^2 \rho^2 \rho'}{(c^2-1)^2}
  - i \kappa f
\right],
\end{align}
where $L$ is the length of the contour in the $x$ direction
and $\kappa$ is given by (\ref{kappa-def}).
The action is now manifestly linear in $M L$, and we also assume that $\kappa $ is kept fixed as indicated in (\ref{scalinglaw-sugra}).

Since the action does not depend on any derivatives of $f$, the equation of motion for $f$ is simply
$\frac{\partial \mathcal{L}}{\partial f} = 0$,
which gives
\begin{align} \label{eq: eomf}
 f =  \dfrac{i}{(c^2-1)^{3/2}}
		     \left(
		      \dfrac{ 1 + A^2  (c^2-1) \rho '^2}{1 + A^2 \dfrac{ c^2 \rho^4}{(c^2-1)^2 \kappa ^2}}
		     \right)^{1/2}.
\end{align}
Integrating out $f$ from the action, we get
\begin{align}\label{eq: D3actionNof}
S = \frac{2 N M L}{\pi} \int dc \left[
  \sqrt{
    \left(
      \kappa^2 + \left(\frac{c A \rho^2}{c^2 - 1}\right)^2
    \right)
    \left(
      \frac{A^2 \rho'^2}{(c^2-1)^2} + \frac{1}{(c^2 - 1)^3}
    \right)
      }
  -\frac{c A^2 \rho^2 \rho'}{(c^2-1)^2} \right],
\end{align}
from which we can derive the equation of motion for $\rho$.
Nonetheless, this is a second order differential equation that is hard to solve.
Instead, we will use the supersymmetry-preserving condition for the D-brane,
which will give a first order equation for $\rho$.
Let us proceed on this computation in the next subsection.

\subsection{Supersymmetry of the D3 brane solution}

In general, the background supersymmetry preserved by D-brane configurations corresponds to Killing spinors $\epsilon$ that are consistent with
\footnote{The sign can be positive in other conventions.}
\begin{align}\label{eq: susyCondition}
\Gamma \epsilon =  -\epsilon,
\end{align}
where $\Gamma$ is the kappa symmetry projector for a given D-brane.
The equation (\ref{eq: susyCondition}) is the supersymmetry condition, from which we will derive the constraint necessary to solve the equation of motion for $\rho$.
The Killing spinor for the Pilch-Warner background were found in \cite{Pilch:2003jg}, and in appendix \ref{app:KS} we review the analysis in detail and provide all the required expressions and definitions for the present paper.

\subsubsection{Kappa symmetry projector}

The kappa symmetry projector for a given D-brane configuration in Minkowski signature is given by \cite{Skenderis:2002vf}
\begin{align}
d^{p+1} \xi \Gamma = - e^{-\Phi} L_{\text{DBI}}^{-1} e^{\mathcal{F}}\wedge X|_{\text{Vol}},
\end{align}
where
\begin{align}
L_{\text{DBI}}= e^{-\Phi} \sqrt{-\det  (g+\mathcal{F})}\quad;
\quad
X = \bigoplus_n \gamma_{(2n)} K^n I,
\end{align}
and $|_{\text{Vol}}$ indicates projection to the appropriate volume form as in the LHS.
The operators $K$ and $I$ act on a spinor $\psi$ as $K \psi = \psi^*$ and $I \psi = -i \psi$.
We also defined
\begin{align}
\gamma_{(n)} = \frac{1}{n !}d\xi^{i_n}\wedge ... \wedge d\xi^{i_1} \tilde{\gamma}_{i_1...i_n},
\end{align}
and
\begin{align}
\tilde{\gamma}_{i_1...i_n} = \p_{i_1} X^{\mu_1}...\p_{i_n} X^{\mu_n}\gamma_{\mu_1...\mu_n}.
\end{align}
The projector is traceless and satisfies $\Gamma^2 = 1$.

 \subsubsection{The supersymmetric solution }\label{subsec: solveSUSY}

The kappa symmetry projector for the D3-brane, using the induced metric (\ref{eq: inducedMetric}), is
\begin{align}
\Gamma
& = - e^{-\Phi} L_{\text{DBI}}^{-1}
\left(
\gamma_{1234} \rho'(c)+\gamma_{1345}
+ \gamma_{23} \dfrac{2\pi}{\sqrt{\lambda}} F_{x c}
K
\right)I,
\end{align}
where the $\gamma$s are in the curved target space
\footnote{We use $x^\mu=\{x_1,\rho,\omega,\eta,c,...\}$, where $\omega$ and $\eta$ parametrize the 2-sphere. }.
We shall use now $\gamma_\mu = e^a_\mu \Gamma_a$, where $\Gamma_a$ are the constant Dirac matrices and $e^a_\mu$ are the vielbeins of the target space (\ref{eq: deformedAdS5}),
i.e. $G_{\mu\nu } = e^a_\mu e^b_\nu \eta_{a b}$.
Since our metric is diagonal, the projector can be written as,
\begin{align}\label{eq: kappaprojeq}
\Gamma
& = - e^{-\Phi} L_{\text{DBI}}^{-1} \rho'(c) e_1^1 e_2^2 e_3^3 e_4^4 \; \Gamma_{1234}
\left(
1 + \dfrac{e_5^5}{ e_2^2}\Gamma_{25}
+ \dfrac{1}{e_1^1 e_4^4}\Gamma_{23} \dfrac{2\pi}{\sqrt{\lambda}} F_{x c} K
\right)I.
\end{align}

Let us rescale $\rho \rightarrow \rho/M$ and use $f=\dfrac{2\pi}{\sqrt{\lambda} M} F_{x c}$ as we did for the action.
Then, the projector written explicitly is
\begin{align}\label{eq: kappaprojExplicit}
\Gamma & = -
\left(1+\frac{1 -f^2 (c^2-1)^3}{(c^2-1) A^2 \rho'^2}\right)^{-1/2}\Gamma_{1234}
\left(
1
-\frac{1}{ A \rho' \sqrt{c^2-1} }\Gamma_{2 5}
  \left(
      1 + f \left(c^2-1\right)^{3/2}  \Gamma_{1 5} K
  \right)
\right)I.
\end{align}

This projector applies to the Killing spinor (\ref{eq: KillingSpinorClassic})
and it must satisfy the supersymmetric condition (\ref{eq: susyCondition}).
Using,
\begin{align}
\Gamma_{1234}I\epsilon = i \Gamma_{1234}\epsilon = \epsilon,
\end{align}
which is derived from $i\Gamma_{1234}\Pi_-^0 = \Pi_-^0$ (see appendix \ref{app:KS}),
we see immediately that the global prefactor in (\ref{eq: kappaprojExplicit}) should be equal to 1,
which is also the same condition for the prefactor in front of $\Gamma_{1 5}$ to be $\pm 1$.
We choose the minus sign, because the constraint
\begin{align}
 f = - \dfrac{1}{(c^2-1)^{3/2}},
\end{align}
is compatible with the equation of motion for $f$ (compare with (\ref{eq: eomf}) multiplied by $i$ since we are in the Minkowski signature now).
From the comparison, we also obtain a condition for $\rho'^2$, where the positive choice
\begin{align}\label{eq: solRho}
 \rho' =  \dfrac{c \rho^2}{\kappa} (c^2-1)^{-3/2},
\end{align}
is the one that solves the equation of motion for $\rho$ derived from the action (\ref{eq: D3actionNof}).

The kappa projector (\ref{eq: kappaprojeq}) at the solution is then,
\begin{align}\label{eq: kappaprojeq2}
\Gamma
& = -
\Gamma_{1234}\left(
1
-\frac{\kappa (c^2-1) }{A c \rho^2   }\Gamma_{25}
\left(
  1-\Gamma_{15}K\right)
\right)I,
\end{align}
and the supersymmetry condition is satisfied provided that,
\begin{align}
\frac{1}{2}\left(1-\Gamma_{15}K\right)\epsilon = 0,
\end{align}
i.e. the D3-brane breaks exactly half of the supersymmetries.

The general solution to the differential equation (\ref{eq: solRho}) is,
\begin{align}
\rho(c)=\frac{\kappa \, \sqrt{c^2-1} }{1 - \alpha \, \sqrt{c^2-1} }
\end{align}
where $\alpha$ is an integration constant. 
We fix $\alpha=0$ in order for the solution to coincide with the known $\AdS_5 \times \Sphere^5$ solution of \cite{Drukker:2005kx} close to the boundary ($c\simeq 1$).
Substituting $c=1+z^2/2$ and expanding around $z=0$ we get
\begin{align}
 \rho = \kappa z + \mathcal{O}(z^3).
\end{align}
In section \ref{subsec: newSols} we will discuss the general solution in more detail.

\subsection{The D3-brane action at the solution}

Finally, let us compute the action at the solution.
Plugging (\ref{eq: solRho}) into the action (\ref{eq: D3actionNof}), the DBI and the WZ terms cancel each other, and we are left only with the Lagrange multiplier term
\begin{align}
S =\frac{2 N M L}{\pi} \int_{1+\epsilon^2/2 }^{\infty} dc \: \frac{\kappa}{(c^2-1)^{3/2}} = \frac{2 N M L}{\pi} \kappa \left(\frac{1}{\epsilon} - 1  \right).
\end{align}
Here, $\epsilon$ is the cutoff for the radial coordinate $z$ of the undeformed AdS, since $c=1+z^2/2$ close to the boundary.
Dropping the perimeter divergence, we obtain the finite renormalized action.
Using $\kappa = \sqrt{\lambda}k / (4N)$, the result matches with the matrix model result (\ref{eq: MatrixModelResult}):
\begin{align}
S_{ren} = - \frac{\sqrt{\lambda} k M L}{2\pi}.
\end{align}

\section{Other D3-brane solutions}\label{subsec: newSols}

As we showed in the previous section, there is a family of supersymmetric D3-brane solutions, given by
\begin{align}
\rho(c)=\frac{\kappa \, \sqrt{c^2-1} }{1 - \alpha \, \sqrt{c^2-1} }.
\end{align}
The same also applies for the $\AdS_5 \times \Sphere^5$ case, for
\begin{align}
\rho = \frac{\kappa \, z }{1 - \alpha \, z}.
\end{align}
The main difference between these two cases is that the $\AdS_5 \times \Sphere^5$ background has conformal symmetry.
One can use this symmetry to rescale $z\to \lambda z$ and $\rho\to \lambda \rho$ in order to get
\begin{align}
z = \frac{\rho}{\kappa - \lambda \alpha \rho}.
\end{align}
So in principle there are three distinct cases, $\alpha=0,\pm 1$, while in the PW case different $\alpha$ solution are not related, so we have a one-parameter family of solutions.
Because of the conformal symmetry, it is easier to study the $\AdS_5 \times \Sphere^5$ as we shall do next.
This is enough for our consideration since close to the boundary the solutions should coincide.

\begin{figure}
    \centering
    \includegraphics[trim = 0mm 0mm 0mm 0mm,clip,width = 0.4\textwidth]{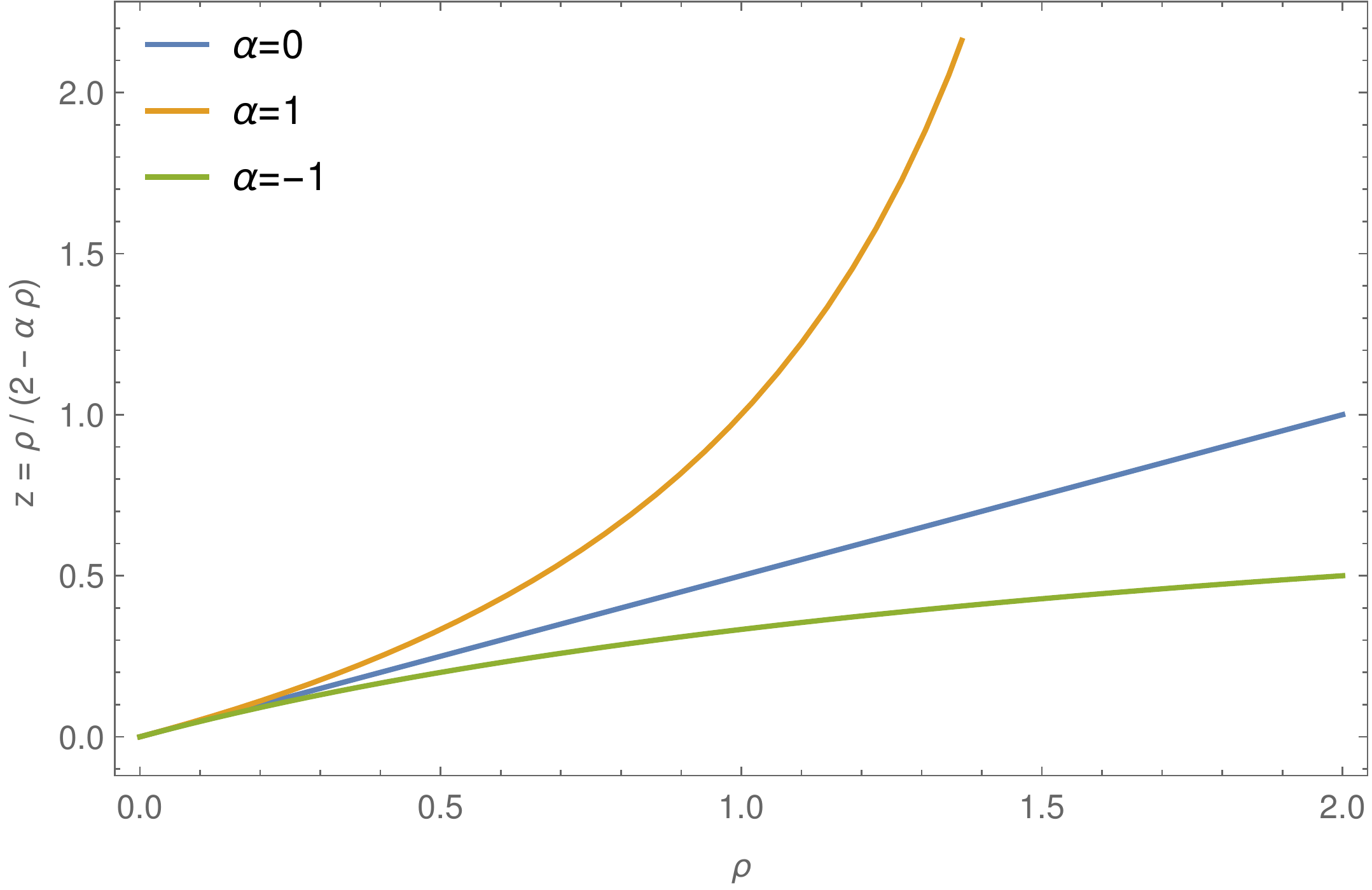}
    \caption{The graph shows the function $z = \frac{\rho}{\kappa - \alpha \rho}$ for the probe D3-brane in the $\AdS_5 \times \Sphere^5$ background for $\kappa=2$ and $\alpha = +1,0,-1$, represented by the orange, blue and green lines respectively.}
    \label{fig:D3braneprofAdS}
\end{figure}

First, we note that inserting the general solution to the action results in
\begin{align}
S_{D3} = \frac{2 N}{\pi}\int \frac{dz}{z^2}.
\end{align}
Thus, the $\alpha$ dependence can appear only through the integration limits.
Indeed, asymptotically the solutions have different behaviors as can be seen in figure \ref{fig:D3braneprofAdS}.
For $\alpha=+1$, the two-sphere's radius grows with $z$, but stays finite ($\rho = \kappa$) when $z\to\infty$.
For $\alpha=0$, the two-sphere's radius grows linearly with $z$.
Finally, for $\alpha=-1$ the Wilson loop solution does not go to infinity in the $z$ direction, and the radius of the two-sphere goes to infinity for a finite value of $z=1$, thus we discard this solution.
So we see that for $\alpha=+1,0$ the action looks the same and has the same limits of integration, and both yield the same result.

\begin{figure}
\begin{center}
\subfloat[][]{
  \includegraphics[width=50mm]{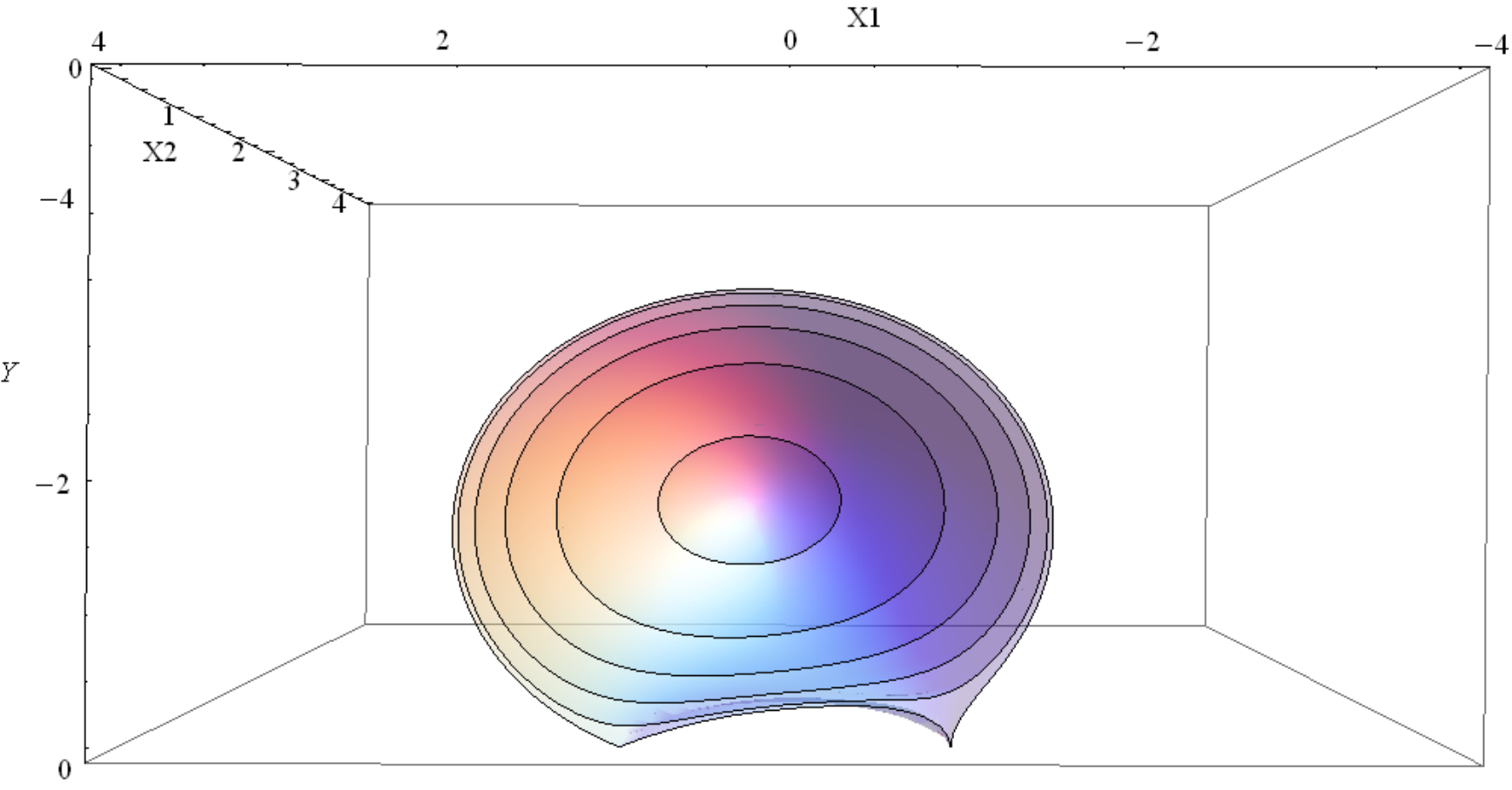}} \hspace{3mm}
\subfloat[][]{
  \includegraphics[width=50mm]{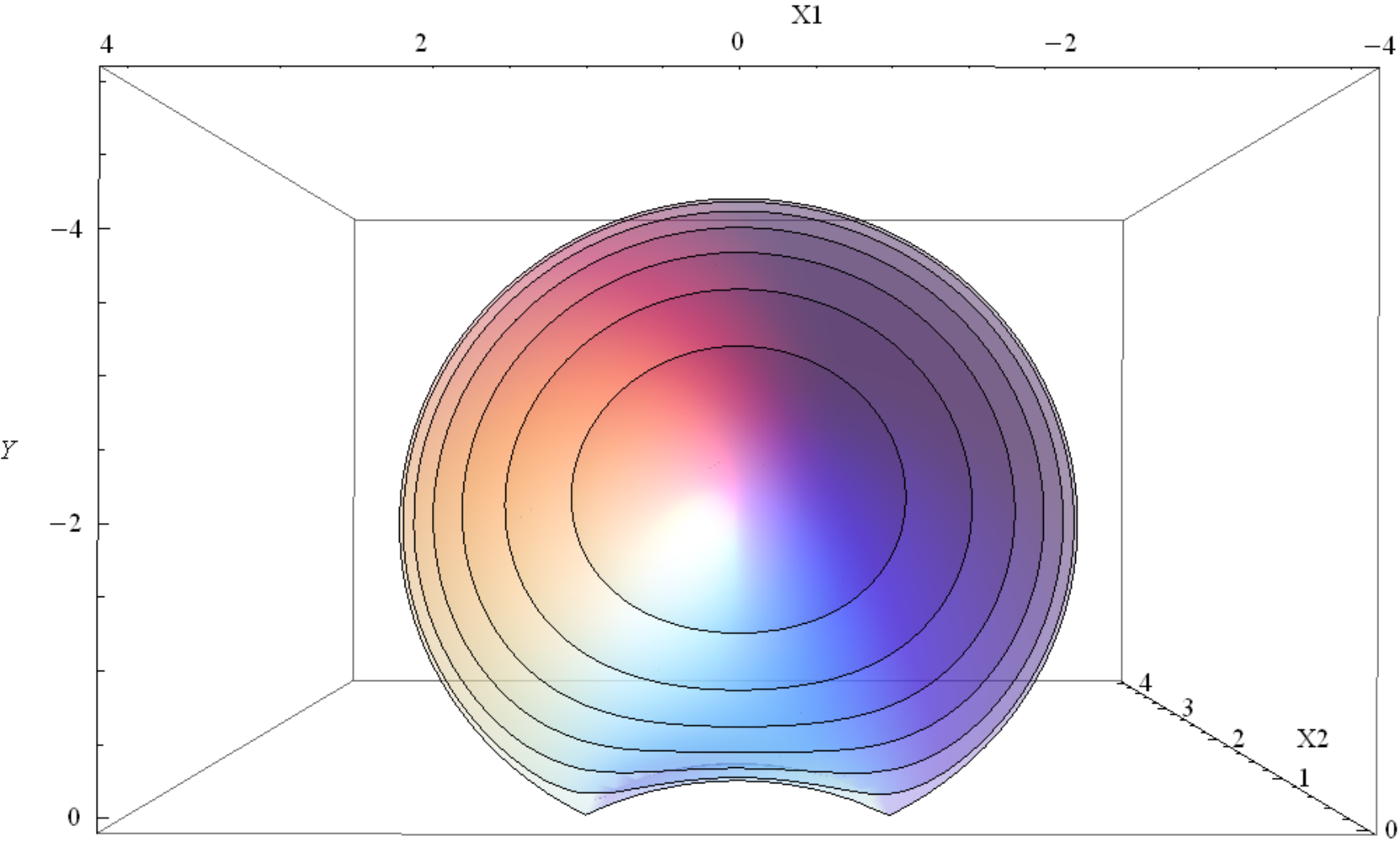}} \hspace{3mm}
\subfloat[][]{
  \includegraphics[width=50mm]{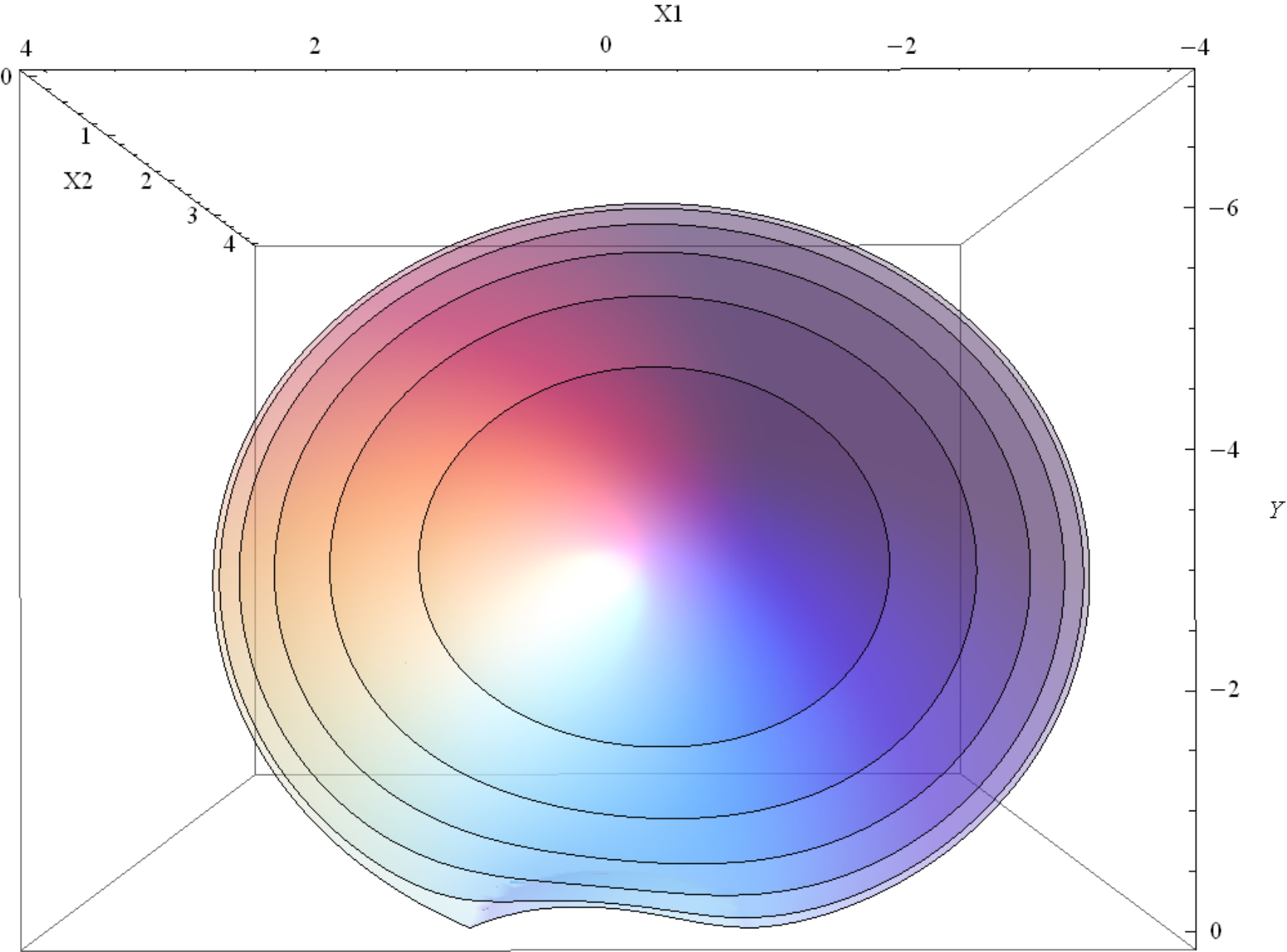}} \hspace{3mm}
\caption{D3-brane solution in $\AdS_5 \times \Sphere^5$ for $\kappa=2$. Parameterizing the metric using $ds^2 = \frac{dx_1^2 + dx_2^2 + dr^2 +r^2d\Omega^2_2 + dz^2}{z^2}$,
the surface depends on  $x_1, x_2, r, z$, so we plot the $x_1, x_2, z$ subspace for different values of $r$.
$r=0$ corresponds to the outer surfaces which approach the AdS boundary, as $r$ grows we get the inner surfaces until a critical value where the surface becomes a point. The different figures correspond to  (a) $\alpha=1$, (b) $\alpha=0$, (c) $\alpha=-2$. As one can see, taking $a\neq 0$ breaks the rotational symmetry of the surface in the $x_1 - x_2$ plane.}
\label{fig:D3braneAdS3D}
\end{center}
\end{figure}

The existence of extra solutions (even in the $AdS_5\times S^5$ case) is rather surprising, and we would like to understand which of these solutions gives the correct
dual description of the Wilson loop. Clearly, the solution with $\alpha =0$ is special, and we are going to motivate this choice by the following considerations.
Suppose that we apply a conformal inversion to the solution with arbitrary $\alpha $. When the inversion maps the Wilson line again to an infinite line, the solution with $\alpha =0$ stays the same, while the $\alpha=+1$ solution is no longer symmetric under translation along the $x_1$ axis, but develops a "bump".
Similarly, we could transform the line to the circle and see a similar picture, where the $\alpha\neq 0$ solutions do not have rotational symmetry in the bulk as one expects for the circular Wilson loop solution, see figure \ref{fig:D3braneAdS3D}.
Thus, only the $\alpha=0$ solution preserves all the symmetries of the problem.
One more difference between the solutions is their topology. The $\alpha=0$ solution is known to have the topology of $\AdS_2 \times \Sphere^2$, while the $\alpha >0 $ case has $\AdS_2 \times \Sphere^2$ topology only close to the boundary, while away from the boundary it is no longer a product space, and in the limit $z\to \infty$ it looks like $\AdS_4$.

\section{Conclusions}

In this paper we constructed the D3-brane embedding in the Pilch-Warner geometry that describes straight Wilson line in the symmetric representations of $SU(N)$. The expectation value inferred from the action of the D3-brane perfectly matches with the localization predictions \cite{Chen-Lin:2015dfa}. The field-theory calculation of \cite{Chen-Lin:2015dfa} actually applies to a more general scaling limit (\ref{scalinglaw-SYM}) compared to the supergravity calculation, allowing the rank of representation to grow with the length of the contour. On the supergravity side, this would correspond to the world-volume field strength becoming an extensive quantity. We do not know how to analyze this regime holographically, which would be very interesting to do, because this regime should reveal the non-trivial phase structure of \sstar at strong coupling.

Near the boundary, our solution reduces to the know D3-brane embedding in $AdS_5$, dual to symmetric-representation Wilson line in $\mathcal{N}=4$ SYM \cite{Drukker:2005kx}. It would be interesting to find a counterpart of the D5-brane solution \cite{Yamaguchi:2006tq}, dual to the Wilson lines in the antisymmetric representations. The localization predictions are available in this case as well \cite{Chen-Lin:2015dfa}, and are better consistent with the scaling of parameters expected from supergravity.

\subsection*{Acknowledgments}

We would like to thank N. Drukker and D. Medina Rinc\'{o}n for useful discussions.
This work was supported by the Marie
Curie network GATIS of the European Union's FP7 Programme under REA Grant
Agreement No 317089, by the ERC advanced grant No 341222. The work of K.Z. was supported in addition by the Swedish Research Council (VR) grant
2013-4329, and by RFBR grant 15-01-99504.

\appendix

\section{Review of the Pilch-Warner geometry}\label{sec:PWB}
First of all, we invite the readers to check out the online repository:
\url{https://github.com/yixinyi/PilchWarner},
where Mathematica packages and notebooks related to this project are hosted. 

In this section we shall describe the Pilch-Warner solution to the type IIB supergravity equations, which are specified in appendix \ref{app:SUGRAEQ}.
We start by introducing the various background fields which are all non-trivial for this solution.
Our notation and conventions are summarized in appendix \ref{app:notation}.
Afterwards, we introduce the Killing spinors for the background, first given in in \cite{Pilch:2003jg}, which are necessary for the supersymmetry analysis of the D-brane solution.

\subsection{Background fields}
In this subsection we quote the expressions for the field content of the Pilch-Warner solution.
We follow mostly the conventions of \cite{Pilch:2003jg} and \cite{Buchel:2000cn}.
\begin{itemize}
  \item \textbf{The metric}\\
The metric (in the Einstein frame) is given by
\begin{align}\label{eq:PWmetric}
ds_E^2 =
\Omega^2 dx_\mu dx^\mu -\left(
V_r^2 dr^2 + V_\theta^2 d\theta^2 + V_1^2 \sigma_1^2 + V_{23}^2 (\sigma_2^2 + \sigma_3^2) + V_\phi^2 d\phi^2\right),
\end{align}
where we use the mostly minus convention.
The various coefficients are given by
\begin{align}
& \Omega = \frac{c^{1/8} A^{1/4} X_1^{1/8} X_2^{1/8}}{(c^2 - 1)^{1/2}},\quad
V_r = \frac{c^{1/8}X_1^{1/8} X_2^{1/8}}{A^{1/12}},\quad
V_\theta = \frac{X_1^{1/8} X_2^{1/8}}{c^{3/8}A^{1/4}},\nonumber\\
&
V_1 = \frac{A^{1/4}X_1^{1/8} }{c^{3/8}X_2^{3/8}},\quad
V_{23} = \frac{c^{1/8}A^{1/4}X_2^{1/8} }{X_1^{3/8}},\quad
V_\phi = \frac{c^{1/8}X_1^{1/8} }{A^{1/4}X_2^{3/8}},
\end{align}
and
\begin{align}
X_1 = & \cos^2\theta + cA  \sin^2\theta,\nonumber\\
X_2 = & c \cos^2\theta + A  \sin^2\theta.
\end{align}
$c$ and $A$ are functions of the radial holographic coordinate $r$ which satisfy
\begin{align}
A  = c+(c^2 - 1)\frac{1}{2}\ln\left(\frac{c-1}{c+1}\right),
\end{align}
and
\begin{align}\label{eq: dcdr}
\frac{dc}{dr} = A^{2/3}(1-c^2),\quad
\frac{d A}{dr} = 2 A^{2/3}\left(1 - c A\right).
\end{align}
Finally, $\sigma_i(\alpha,\beta,\psi) = \tr(g^{-1}\tau_i d g)$ are the $SU(2)$ left invariant forms parameterizing $S^3$, where $\tau_i$ are the Pauli matrices. The one forms satisfy the relation\footnote{Notice that this convention is different than the one used for example in \cite{Buchel:2000cn}, where $d\sigma_i  = -\epsilon_{i j k} \sigma_j \wedge \sigma_k$.} $d\sigma_i  = \epsilon_{i j k} \sigma_j \wedge \sigma_k$.
For example, one can parameterize these one forms using the Euler angles of $S^3$
\begin{align}
\sigma_1 = &\frac{1}{2}\left(d\alpha +\cos\psi \, d\beta\right),\nonumber\\
\sigma_2 = &\frac{1}{2}\left(\cos\alpha \,d\psi +\sin\alpha \sin\psi \, d\beta\right),\nonumber\\
\sigma_3 = &\frac{1}{2}\left(\sin\alpha \, d\psi -\cos\alpha \sin\psi \, d\beta\right),
\end{align}
with $0 \leq \alpha \leq 2\pi$, $0 \leq \beta \leq 2\pi$, $0 \leq \psi \leq \pi$.
The deformed five-sphere part of the metric has an $SU(2)\times U(1)^2$ symmetry, where the $U(1)$'s correspond to $\phi$ translation and a rotation between $\sigma_1$ and $\sigma_2$.

The boundary of the Pilch-Warner geometry is located at $c = 1$.
Defining the coordinate $c = 1 + \frac{z^2}{2}$, and expanding around $z = 0$, we get the $\AdS_5$ metric
\begin{align}\label{eq:metricatc1}
ds_E^2 =
\frac{dx^2 - dz^2}{z^2} + O(z^{0}).
\end{align}
Thus, close to the boundary, $z$ plays the role of the familiar radial AdS coordinate in Poincare coordinates.
Similarly, the rest of the metric reduces to the $S^5$ metric,
\begin{align}
ds_{S^5}^2 =
-\left(d\theta^2 + \sin^2\theta d\phi^2 + \cos^2\theta\left(\sigma_1^2 + \sigma_2^2 + \sigma_3^2\right)\right) + O(z^{1}).
\end{align}

  \item \textbf{Dilaton-Axion field}\\
The field is given by
\begin{align}
V = f\left(
  \begin{array}{cc}
    1 & B \\
    B^* & 1 \\
  \end{array}
\right),
\end{align}
where
\begin{align}
f = \cosh\left(\frac{1}{4}\ln\left(\frac{c X_1}{X_2}\right)\right),\quad
f B = e^{2 i \phi}\sinh\left(\frac{1}{4}\ln\left(\frac{c X_1}{X_2}\right)\right).
\end{align}

Using the relation $C_{(0)} + i e^{-\Phi} = i\frac{B+1}{B-1}$
we get the explicit relation to the RR zero form and the dilaton
\begin{align}
C_0 = \frac{2 b \sin 2\phi}{1-2 b \cos 2\phi + b^2},\quad
e^{-\Phi} = \frac{b^2 - 1}{1-2 b \cos 2\phi + b^2},
\end{align}
where $b = \tanh\left(\frac{1}{4}\ln\left(\frac{c X_1}{X_2}\right)\right)$.

  \item \textbf{Three form}\\
  The antisymmetric three form is given by $F_{(3)} = d A_{(2)} = d\left(C_{(2)} + i B_{(2)}\right)$ where \cite{Pilch:2000ue,Pilch:2003jg}
\begin{align}
A_{(2)} = & e^{i \phi}\left(a_1 d\theta \wedge \sigma_1 + a_2 \sigma_2 \wedge \sigma_3 + a_3 \sigma_1 \wedge d\phi\right),\nonumber\\
a_1 = & - i \frac{\sqrt{c^2-1}}{c}\cos\theta,\nonumber\\
a_2 =  & i A  \frac{\sqrt{c^2-1}}{X_1}\sin \theta \cos^2 \theta,\nonumber\\
a_3 =  &  -\frac{\sqrt{c^2-1}}{X_2}\sin \theta \cos^2 \theta.
\end{align}

  \item \textbf{Five form}\\
  The five form is given by $F_{(5)} = \mathcal{F} + \ast \mathcal{F}$, where\footnote{Notice that we use $c^{1/2}$ in the denominator in contrast to $c$ which appears in \cite{Pilch:2003jg} without the square root.}
\begin{align}
\mathcal{F} = & dx^0 \wedge dx^1 \wedge dx^2 \wedge dx^3 \wedge dw,\nonumber\\
w = & \frac{\Omega^4 X_1^{1/2}}{4 c ^{1/2} X_2^{1/2}    }.
\end{align}

\end{itemize}

\section{Notation}\label{app:notation}
Since the supergravity solution is quite involved and different conventions appear in the literature,
it is worthwhile to summarize the conventions we shall use in this paper.

\subsection{Metric}
The metric will be denoted by $G_{\mu \nu}$ where $\mu,\nu = 1,...,10$, and we use the mostly minus convention.
We will use the indices $a,b=1,...,10$ to denote the metric and other fields components in the non-coordinate basis, related to the curved space by the use of the vielbeins $e_\mu^a$ and their inverse $E_a^\mu$, i.e.
\begin{equation}
G_{\mu \nu} = e^a_\mu e^b_\nu \eta_{ab},\quad
E^{\mu a} = G^{\mu \nu }e_\nu^a.
\end{equation}

\subsection{Background fields}
Following \cite{Buchel:2000cn}, we denote by $\Phi$, $B_{(2)}$ and $C_{(n)}$ the dilaton, 2-form NSNS field and the RR potentials respectively.
We further define
\begin{align}\label{eq:defs1}
&C_{(0)} + i e^{-\Phi} = i\frac{B+1}{B-1},\nonumber\\
&A_{(2)} = C_{(2)} + i B_{(2)},\nonumber\\
&A_{(4)} = \frac{1}{4}\left(C_{(4)} + \frac{1}{2}B_{(2)}\wedge C_{(2)}\right),\nonumber\\
&F_{(3)} = d A_{(2)},\nonumber\\
&F_{(5)} = d A_{(4)} - \frac{1}{8} \Im \left(A_{(2)}\wedge F_{(3)}^*\right). 
\end{align}
We also have
\begin{align}\label{eq:defs2}
&\tilde F_{(1)} = d C_{(0)},\nonumber\\
&\tilde F_{(3)} = d C_{(2)} + C_{(0)} d B_{(2)},\nonumber\\
&\tilde F_{(5)} = d C_{(4)} + C_{(2)} \wedge d B_{(2)},\nonumber\\
&\tilde F_{(7)} = d C_{(6)} + C_{(4)} \wedge d B_{(2)},\nonumber\\
&\tilde F_{(9)} = d C_{(8)} + C_{(6)} \wedge d B_{(2)},\nonumber\\
\end{align}
and the duality relation
\begin{align}\label{eq:dualityconstraint}
\ast \tilde F_{(n+1)} =(-)^{n(n-1)/2} \tilde F_{(9-n)}.
\end{align}
Notice that by these definitions $\tilde F_{(5)} = 4 F_{(5)}$.

Finally, the supergravity and Killing spinor equations involve the following filed combinations
\begin{align}\label{eq:defs3}
P_\mu & = f^2 \p_\mu B,\nonumber\\
Q_\mu & = f^2 \Im(B \p_\mu B^*),\nonumber\\
G_{\mu \nu \rho} & = f\left(F_{\mu \nu \rho} - B F_{\mu \nu \rho}^*\right).
\end{align}

\subsection{p-Forms}
We use the standard differential forms notation where
\begin{align}
& A_p = \frac{1}{p!} A_{\mu_1..\mu_p}d x^{\mu_1}\wedge ..\wedge d x^{\mu_p} = \frac{1}{p!} A_{[\mu_1..\mu_p]}d x^{\mu_1}\wedge ..\wedge d x^{\mu_p},
\end{align}
where the square brackets indicate antisymmetrization with the proper normalization.
Notice for example that if $F_{(n+1)} = d C_{(n)}$, then $F_{\mu_1 .. \mu_{n+1}} = (n+1) \p_{[\mu_1} C_{\mu_2 .. \mu_{n+1}]}$.

We transforms between the target space and pullback indices using
\begin{align}
A_{\mu_1..\mu_n a_1 .. a_m}
=
A_{\mu_1..\mu_n \nu_1 .. \nu_m} \p_{a_1} X^{\nu_1}..\p_{a_m} X^{\nu_m}.
\end{align}
\section{Supergravity equations}\label{app:SUGRAEQ}
The supergravity equations are given by \cite{Schwarz:1983qr} (the various field definitions are found in appendix \ref{app:notation})
\paragraph{Einstein equation}
\begin{align}
R_{\mu \nu} = T^{(1)}_{\mu \nu} + T^{(3)}_{\mu \nu} + T^{(5)}_{\mu \nu},
\end{align}
where
\begin{align}
T^{(1)}_{\mu \nu} & = P_\mu P_\nu^* + P_\nu P_\mu^*,\nonumber\\
T^{(3)}_{\mu \nu} & = \frac{1}{8}\left(
  G^{\sigma\rho}{}_{\mu} G_{\sigma\rho\nu}^*
+ G^{\sigma\rho}{}_{\nu} G_{\sigma\rho\mu}^*
- \frac{1}{6}\eta_{\mu\nu}G_{\sigma\rho\chi}^* G^{\sigma\rho\chi}\right),\nonumber\\
T^{(5)}_{\mu \nu} & = \frac{1}{6}F^{(5)}{}_{\rho\sigma\chi\zeta\mu}F^{(5)}{}^{\rho\sigma\chi\zeta}{}_{\nu}.
\end{align}
\paragraph{Maxwell equation}
\begin{align}
(\nabla_\lambda -i Q_\lambda)G^{\mu \nu \lambda} - P_\lambda (G^{\mu \nu \lambda})^* + i \frac{2}{3}F^{(5)}{}^{\mu\nu\lambda\rho\sigma}G_{\lambda\rho\sigma} = 0.
\end{align}
\paragraph{Dilaton equation}
\begin{align}
(\nabla_\mu -2 i Q_\mu)P^\mu  + \frac{1}{24}G_{\sigma\rho\chi} G^{\sigma\rho\chi} = 0
\end{align}
\paragraph{Self-dual equation}
\begin{align}
F^{(5)} = \ast F^{(5)}.
\end{align}

\section{Killing spinors in the Pilch-Warner background}\label{app:KS}

The Killing spinors are defined using the vanishing of the supersymmetry transformation of the gravitino and dilatino, given by
\begin{align}\label{eq:KSE}
& \delta \psi_\mu = D_\mu \epsilon + \frac{i}{480} F_{\rho_1 \rho_2 \rho_3 \rho_4 \rho_5}\gamma^{\rho_1 \rho_2 \rho_3 \rho_4 \rho_5} \gamma_\mu \epsilon + \frac{1}{96}\left(\gamma_\mu{}^{\nu \rho \lambda}G_{\nu \rho \lambda} - 9 \gamma^{\rho \lambda} G_{\mu \rho \lambda}\right)\epsilon^*,\nonumber\\
& \delta \lambda = i P_\mu \gamma^\mu \epsilon^* - \frac{i}{24}G_{\mu \nu \rho}\gamma^{\mu \nu \rho}\epsilon,
\end{align}
where $\epsilon$ is a chiral spinor $\Gamma^{11}  \epsilon = - \epsilon$, ($\Gamma^{11}\equiv \Gamma^1 \Gamma^2..\Gamma^{10}$), and
\begin{align}
D_\mu \epsilon  = \p_\mu \epsilon +\frac{1}{4}\omega_{\mu}{}^{ab}\Gamma_{ab}\epsilon -\frac{i}{2}Q_\mu \epsilon,
\end{align}
where the spin connection is defined as follows
\begin{align}
&\omega_{\mu \nu \rho} = -( \Omega_{\nu \rho \mu } + \Omega_{\nu \mu \rho} + \Omega_{\mu \rho \nu } )\nonumber\\
&\Omega_{\mu \nu \rho} = e^a{}_\rho \p_{[\mu}e_{\nu] a}. \nonumber
\end{align}
Throughout this section we use the metric (\ref{eq:PWmetric}).
Gamma matrices with Latin indices are constant matrices and with Greek indices are generally not, given by $\gamma_\mu = e_\mu^a \Gamma_a$.

The Killing spinors for the PW background were found in \cite{Pilch:2003jg}.
In this section we shall review the analysis, using the assumptions given in \cite{Pilch:2003jg}, namely starting with the ansatz
\begin{align}
\epsilon = e^{i\phi/2}\mathcal{M}(r,\theta)\epsilon_0,
\end{align}
where $\mathcal{M}(r,\theta)$ is a matrix and $\epsilon_0$ is a spinor which depends only on the $\mathfrak{su}(2)$ coordinates.

First, as in \cite{Pilch:2003jg}, we note that for the following combination of Killing spinor equations,
which is zero by definition,
the $\epsilon^*$ contribution gets canceled out, leading to a projector for $\epsilon$:
\begin{align}
0=2(\gamma^1\delta\psi_1 + \gamma^{10}\delta\psi_{10}) + ie^{2 i \phi}(\delta\lambda)^*
\quad
\Longrightarrow \quad
 \mathcal{P}_+(\alpha) \epsilon =\epsilon,
\end{align}
where
\begin{align}
\mathcal{P}_\pm (\alpha)= \frac{1}{2}\left(
					  1 \pm i
					  \left(
					      \cos\alpha \Gamma^{5,10} + \sin\alpha \Gamma^{6,10}
					  \right)
				      \right)
\end{align}
with
\begin{align}
\cos\alpha \equiv \frac{\cos\theta}{\sqrt{X_1}}, \quad
\sin\alpha \equiv \frac{\sqrt{c}A^{1/2}\sin\theta}{\sqrt{X_1}}.
\end{align}
These projectors also satisfy
\begin{align}\label{eq: proj610}
\mathcal{P}_\pm (\alpha) = O(\alpha) \mathcal{P}_{\pm}^{0}O^{-1}(\alpha), \quad
\mathcal{P}_{\pm}^{0} \equiv \dfrac{1}{2} \left(1\pm i\Gamma^{6,10}\right)
\end{align}
where
\begin{align}\label{eq:rotmat56}
O(\alpha)  = \cos\frac{\alpha}{2} - \sin\frac{\alpha}{2}\Gamma^{5 6},
\quad
O^{-1}(\alpha) = \cos\frac{\alpha}{2} + \sin\frac{\alpha}{2}\Gamma^{5 6} .
\end{align}

Second, the dilatino equation, together with the previous projector condition and the chirality condition,
lead to another projector, i.e.
\begin{align}
0=\delta \lambda, \quad
 \mathcal{P}_+(\alpha)\epsilon =\epsilon, \quad
 \Gamma^{11}  \epsilon = - \epsilon \quad
 \Longrightarrow \quad
\Pi_- (\beta) \epsilon = \epsilon,
\end{align}
where
\begin{align}
\Pi_\pm (\beta)= \frac{1}{2}\left( 1 \pm i\Gamma^{1234} \left(\cos\beta - \sin \beta \Gamma^{7,10} * \right)\right),
\end{align}
with
\begin{align}
\cos\beta \equiv \frac{X_1}{c X_2}, \quad
\sin\beta \equiv -\sqrt{\frac{c^2 - 1}{c X_2}}\cos\theta.
\end{align}
These projectors also satisfy
 \[
 \Pi_\pm(\beta) = O^*(\beta)\,\Pi_\pm^0 \, O^*(\beta)^{-1}, \quad
 \Pi_\pm^0 = \frac{1}{2}\left(1 \pm i\Gamma^{1234}\right).
\]
where
\begin{align}\label{eq:rotmatStar}
O^*(\beta) = \cos\frac{\beta}{2} + \sin\frac{\beta}{2}\Gamma^{7,10}*, \quad
O^*(\beta)^{-1} = \cos\frac{\beta}{2} - \sin\frac{\beta}{2}\Gamma^{7,10}*, \quad
\end{align}

Thus, so far the Killing spinor satisfies
\begin{align}
\Pi_- (\beta) \mathcal{P}_+ (\alpha) \epsilon = \epsilon.
\end{align}

Next, we have to check nine more gravitino equations.
We start with the flat directions where the equations are purely algebraic since we assume $\epsilon$ to be independent of $x^\mu$.
Using the two projectors and chirality condition it is straight forward to show that $\delta \psi_1 = ... = \delta \psi_4 = 0$.
Regarding the $\mathfrak{su}(2)$ directions, we use again the projectors and chirality condition and we get
\begin{align}
& \delta \psi_7 = \left(\p_{\sigma_1} - \Gamma^{89}\right)\epsilon \nonumber \\
& \delta \psi_8 = \left(\p_{\sigma_2} + \left(\cos\alpha\Gamma^{58} - \sin\alpha\Gamma^{68}\right)\right)\epsilon  \nonumber\\
& \delta \psi_9 = \left(\p_{\sigma_3} + \left(\cos\alpha\Gamma^{59} - \sin\alpha\Gamma^{69}\right)\right)\epsilon  \nonumber
\end{align}
where by $\p_{\sigma_i}$ we mean the dual of $\sigma_i$. These equation can be summarized in the following form
\begin{align}
\left(\p_{\sigma_i} - O(\alpha)\,t_i \,O^{-1}(\alpha)\right)\epsilon = 0,
\end{align}
with
\begin{align}
t_1 = \Gamma^{89},\quad
t_2 = -\Gamma^{58},\quad
t_3 = -\Gamma^{59}.
\end{align}
Since $[\Pi_-(\beta) \mathcal{P}_+(\alpha), O(\alpha)\,t_i \,O^{-1}(\alpha)] = 0$, we can consider spinors of the form
\begin{align}
\epsilon \sim \Pi_-(\beta) \mathcal{P}_+(\alpha)  O(\alpha) \epsilon_0,
\end{align}
so that $\epsilon_0$ depends only on the $\mathfrak{su}(2)$ coordinates through
\begin{align}\label{eq:su2depnedenceofKS}
\left(\p_{\sigma_i} - t_i \right)\epsilon_0 = 0.
\end{align}
Using (\ref{eq: proj610}) and $[\Pi_-(\beta) , O(\alpha)] = 0$,
\begin{align}
\epsilon \sim \Pi_-(\beta) \mathcal{P}_+(\alpha)  O(\alpha) \epsilon_0
= O(\alpha) \Pi_-(\beta)  \mathcal{P}_+^{0}  \epsilon_0.
\end{align}

Now we are left with two more equations, $\delta\psi_5 = 0$ and $\delta\psi_6 = 0$.
Defining $\delta\psi_M = (\p_M + \Delta_M) \epsilon$ for $M=5, 6$, we have
\begin{align}\label{eq:deltam}
\Delta_M \epsilon
&=\p_M \left(
\frac{1}{16}\ln\left(\frac{4 X_2^3 c^3}{X_1 A^{2} \cos^8\theta}\right)
+\Gamma^{56} \frac{\alpha}{2}
-i\Gamma^{1234} \frac{1}{2}\ln\tan\frac{\beta}{2} \right)\epsilon.
\end{align}
Let us take an ansatz for $\epsilon$ as
\begin{align}
\epsilon
= e^{i\phi/2}O(\alpha) \Pi_-(\beta) M(r,\theta)  \mathcal{P}_+^{0}  \epsilon_0,
\end{align}
with
\[
 M = \frac{a(r,\theta)}{\sin\frac{\beta}{2}}\Pi^0_+ + \frac{b(r,\theta)}{\cos\frac{\beta}{2}}\Pi^0_-.
\]
Because $\frac{1}{2}\p_M \alpha = \Gamma^{56}  O(\alpha)^{-1} \p_M O(\alpha)$,
the second term in (\ref{eq:deltam}) cancels the $O(\alpha)$ derivative of $\epsilon$. The Killing spinor equations are left with
\begin{align}\label{eq:reducedKSE}
\p_M(\Pi_-(\beta) M(r,\theta)  ) + \p_M \left(
\frac{1}{16}\ln\left(\frac{4 X_2^3 c^3}{X_1 A^{2} \cos^8\theta}\right)
-i\Gamma^{1234} \frac{1}{2}\ln\tan\frac{\beta}{2} \right)\Pi_-(\beta) M(r,\theta) =0.
\end{align}
We can write $\Pi_-(\beta) M(r,\theta)$ as follows
\begin{align}
\Pi_-(\beta) M(r,\theta)
& = O^*(\beta)\Pi_-^0 O^*(\beta)^{-1}\left(\frac{a}{\sin\frac{\beta}{2}}\Pi^0_+ + \frac{b}{\cos\frac{\beta}{2}}\Pi^0_-\right)\nonumber\\
& = O^*(\beta)\left(\cos\frac{\beta}{2}\Pi_-^0  - \sin\frac{\beta}{2}\Gamma^{7,10}*\Pi_+^0 \right)\left(\frac{a}{\sin\frac{\beta}{2}}\Pi^0_+ + \frac{b}{\cos\frac{\beta}{2}}\Pi^0_-\right)\nonumber\\
& = \left(\cos\frac{\beta}{2} + \sin\frac{\beta}{2}\Gamma^{7,10}*\right)
\left(b\Pi_-^0  - a\Gamma^{7,10}*\Pi_+^0 \right)\nonumber\\
& =
\Pi_-^0 \, b\cos\frac{\beta}{2}  +  \Pi_+^0 \, b\sin\frac{\beta}{2}\Gamma^{7,10}*
 + \Pi_+^0 \,a \sin\frac{\beta}{2} - \Pi_-^0 \, a \cos\frac{\beta}{2}\Gamma^{7,10}*.
\end{align}
Notice that $i\Gamma^{1234}\Pi_\pm^0 = \pm \Pi_\pm^0$.
Each term should satisfy (\ref{eq:reducedKSE}) independently. It is also easy to see that $a$ and $b$ satisfy the same first order linear differential equation,
which for $b$ is:
\begin{align}
\p_M\left(b\cos\frac{\beta}{2}\right) + \p_M \left(
\frac{1}{16}\ln\left(\frac{4 X_2^3 c^3}{X_1 A^{2} \cos^8\theta}\right)
+\frac{1}{2}\ln\tan\frac{\beta}{2} \right)b\cos\frac{\beta}{2} = 0
\end{align}
so we have
\begin{align}
\p_M\ln \left(b\left(\frac{X_2^3 c^3}{X_1 A^{2} \cos^8\theta}\right)^{1/16}\sin^{1/2}\beta \right) = 0,
\end{align}
and it is solved by
\begin{align}
b = b_0\left(\frac{X_2^3 c^3 \sin^{8}\beta}{X_1 A^{2} \cos^8\theta}\right)^{-1/16} .
\end{align}
Thus, we find that
\begin{align}
\Pi_-(\beta) M(r,\theta)
& = \left(\frac{X_2^3 c^3 \sin^{8}\beta}{X_1 A^{2} \cos^8\theta}\right)^{-1/16} O^*(\beta)
\Pi_-^0\left(b_0  - a_0\Gamma^{7,10}* \right).
\end{align}
In the AdS limit ($c\to 1 + \frac{z^2}{2}$, $z\to 0$) we get
\begin{align}
\Pi_-(\beta) M(r,\theta) \simeq \frac{1}{\sqrt{z}}\Pi_-^0\left(b_0  - a_0\Gamma^{7,10}* \right),
\end{align}
so we choose $a_0 = 0$, and the Killing spinor becomes
\begin{align}
\epsilon = e^{i\phi/2}\Omega^{1/2} O(\alpha)O^*(\beta)
\Pi_-^0 \mathcal{P}_+^{0} \epsilon_0,
\end{align}
where
\[
 \Omega(r,\theta) = \left(\frac{X_1 A^{2} \cos^8\theta}{X_2^3 c^3 \sin^{8}\beta}\right)^{1/8} = \frac{c^{1/8}X_1^{1/8}X_2^{1/8} A^{1/4} }{(c^2-1)^{1/2}}
\]
is the vielbein of the $M=1,..,4$ coordinates.
The solution can be written as the result from \cite{Pilch:2003jg},
\begin{align}
\epsilon = \frac{e^{i\phi/2}\Omega^{1/2}}{\cos\frac{\beta}{2}}O(\alpha)
\Pi_-(\beta) \mathcal{P}_+^{0} \Pi^0_-\epsilon_0, \quad
\mathcal{P}_+^{0} \Pi^0_-\epsilon_0 = \epsilon_0.
\end{align}

In the  $\theta\to\pi/2, \phi \to 0$ limit, which is the regime of our D3-brane configuration,
$\beta = 0$ and $\alpha = \pi/2$ and $\Pi_-(\beta=0) = \Pi_-^0$, hence we find
\begin{align}\label{eq: KillingSpinorClassic}
\epsilon(\theta=\pi/2,\phi=0) = \frac{\sqrt{2}c^{1/8}A^{1/4}}{(c^2-1)^{1/4}}
\frac{1}{2}(1-\Gamma^{56})\mathcal{P}_+^{0} \Pi^0_-\epsilon_0.
\end{align}
Notice that $\frac{1}{2}(1-\Gamma^{56})$ is not a projector.

Last, but not the least, the prescription to use $c$ as coordinate from the original $r$ coordinate is to change $\Gamma^5 \rightarrow -\Gamma^5 $ from the previous result.
This is because $dc/dr<0$, see (\ref{eq: dcdr}), which is a parity transformation.
Thus, the chirality condition of the Killing spinor changes sign. Notice that the hodge dual of the 5-form transforms with a sign too.


\bibliographystyle{nb}

\begin{thebibliography}{10}
\ifx\href\asklfhas\newcommand{\href}[2]{#2}\fi
\ifx\arxivref\asklfhas\newcommand{\arxivref}[1]{\href{http://arxiv.org/abs/#1}{#1}}\fi
\ifx\doiref\asklfhas\newcommand{\doiref}[2]{\href{http://dx.doi.org/#1}{#2}}\fi
\raggedright
\small
\parskip 0pt

\bibitem{Pilch:2000ue}
K.~Pilch and N.~P.~Warner,
\textit{``{N=2 supersymmetric RG flows and the IIB dilaton}''},
\textsf{\doiref{10.1016/S0550-3213(00)00656-8}{Nucl.Phys.~B594,~209~(2001)}},
\texttt{\arxivref{hep-th/0004063}}.
%
\bibitem{Pestun:2007rz}
V.~Pestun,
\textit{``{Localization of gauge theory on a four-sphere and supersymmetric
  Wilson loops}''},
\textsf{\doiref{10.1007/s00220-012-1485-0}{Commun.Math.Phys.~313,~71~(2012)}},
\texttt{\arxivref{0712.2824}}.
%
\bibitem{Buchel:2013id}
A.~Buchel, J.~G.~Russo and K.~Zarembo,
\textit{``{Rigorous Test of Non-conformal Holography: Wilson Loops in N=2*
  Theory}''},
\textsf{\doiref{10.1007/JHEP03(2013)062}{JHEP~1303,~062~(2013)}},
\texttt{\arxivref{1301.1597}}.
%
\bibitem{Bobev:2013cja}
N.~Bobev, H.~Elvang, D.~Z.~Freedman and S.~S.~Pufu,
\textit{``{Holography for $N = 2^*$ on $S^4$}''},
\textsf{\doiref{10.1007/JHEP07(2014)001}{JHEP~1407,~001~(2014)}},
\texttt{\arxivref{1311.1508}}.
%
\bibitem{Chen-Lin:2015dfa}
X.~Chen-Lin and K.~Zarembo,
\textit{``{Higher Rank Wilson Loops in N = 2* Super-Yang-Mills Theory}''},
\textsf{\doiref{10.1007/JHEP03(2015)147}{JHEP~1503,~147~(2015)}},
\texttt{\arxivref{1502.01942}}.
%
\bibitem{Chen:2014vka}
X.~Chen-Lin, J.~Gordon and K.~Zarembo,
\textit{``{$ \mathcal{N}={2}^*$ super-Yang-Mills theory at strong coupling}''},
\textsf{\doiref{10.1007/JHEP11(2014)057}{JHEP~1411,~057~(2014)}},
\texttt{\arxivref{1408.6040}}.
%
\bibitem{Zarembo:2014ooa}
K.~Zarembo,
\textit{``{Strong-Coupling Phases of Planar N=2* Super-Yang-Mills Theory}''},
\textsf{\doiref{10.1007/s11232-014-0232-4}{Theor.Math.Phys.~181,~1522~(2014)}},
\texttt{\arxivref{1410.6114}}.
%
\bibitem{Russo:2013qaa}
J.~G.~Russo and K.~Zarembo,
\textit{``{Evidence for Large-N Phase Transitions in N=2* Theory}''},
\textsf{\doiref{10.1007/JHEP04(2013)065}{JHEP~1304,~065~(2013)}},
\texttt{\arxivref{1302.6968}}.
%
\bibitem{Russo:2013kea}
J.~Russo and K.~Zarembo,
\textit{``{Massive N=2 Gauge Theories at Large N}''},
\textsf{\doiref{10.1007/JHEP11(2013)130}{JHEP~1311,~130~(2013)}},
\texttt{\arxivref{1309.1004}}.
%
\bibitem{Drukker:2005kx}
N.~Drukker and B.~Fiol,
\textit{``{All-genus calculation of Wilson loops using D-branes}''},
\textsf{\doiref{10.1088/1126-6708/2005/02/010}{JHEP~0502,~010~(2005)}},
\texttt{\arxivref{hep-th/0501109}}.
%
\bibitem{Hartnoll:2006hr}
S.~A.~Hartnoll and S.~P.~Kumar,
\textit{``{Multiply wound Polyakov loops at strong coupling}''},
\textsf{\doiref{10.1103/PhysRevD.74.026001}{Phys.~Rev.~D74,~026001~(2006)}},
\texttt{\arxivref{hep-th/0603190}}.
%
\bibitem{Yamaguchi:2006tq}
S.~Yamaguchi,
\textit{``{Wilson loops of anti-symmetric representation and D5-branes}''},
\textsf{\doiref{10.1088/1126-6708/2006/05/037}{JHEP~0605,~037~(2006)}},
\texttt{\arxivref{hep-th/0603208}}.
%
\bibitem{Gomis:2006sb}
J.~Gomis and F.~Passerini,
\textit{``{Holographic Wilson loops}''},
\textsf{JHEP~0608,~074~(2006)},
\texttt{\arxivref{hep-th/0604007}}.
%
\bibitem{Hartnoll:2006is}
S.~A.~Hartnoll and S.~P.~Kumar,
\textit{``{Higher rank Wilson loops from a matrix model}''},
\textsf{JHEP~0608,~026~(2006)},
\texttt{\arxivref{hep-th/0605027}}.
%
\bibitem{RodriguezGomez:2006zz}
D.~Rodriguez-Gomez,
\textit{``{Computing Wilson lines with dielectric branes}''},
\textsf{\doiref{10.1016/j.nuclphysb.2006.06.037}{Nucl.Phys.~B752,~316~(2006)}},
\texttt{\arxivref{hep-th/0604031}}.
%
\bibitem{Gomis:2006im}
J.~Gomis and F.~Passerini,
\textit{``{Wilson loops as D3-branes}''},
\textsf{JHEP~0701,~097~(2007)},
\texttt{\arxivref{hep-th/0612022}}.
%
\bibitem{Yamaguchi:2007ps}
S.~Yamaguchi,
\textit{``{Semi-classical open string corrections and symmetric Wilson
  loops}''},
\textsf{\doiref{10.1088/1126-6708/2007/06/073}{JHEP~0706,~073~(2007)}},
\texttt{\arxivref{hep-th/0701052}}.
%
\bibitem{Fiol:2014vqa}
B.~Fiol, A.~G{\"u}ijosa and J.~F.~Pedraza,
\textit{``{Branes from Light: Embeddings and Energetics for Symmetric
  $k$-Quarks in $\mathcal{N}=4$ SYM}''},
\textsf{\doiref{10.1007/JHEP01(2015)149}{JHEP~1501,~149~(2015)}},
\texttt{\arxivref{1410.0692}}.
%
\bibitem{Faraggi:2011ge}
A.~Faraggi, W.~Mueck and L.~A.~Pando~Zayas,
\textit{``{One-loop Effective Action of the Holographic Antisymmetric Wilson
  Loop}''},
\textsf{\doiref{10.1103/PhysRevD.85.106015}{Phys.~Rev.~D85,~106015~(2012)}},
\texttt{\arxivref{1112.5028}}.
%
\bibitem{Faraggi:2014tna}
A.~Faraggi, J.~T.~Liu, L.~A.~Pando~Zayas and G.~Zhang,
\textit{``{One-loop structure of higher rank Wilson loops in AdS/CFT}''},
\textsf{\doiref{10.1016/j.physletb.2014.11.060}{Phys.Lett.~B740,~218~(2015)}},
\texttt{\arxivref{1409.3187}}.
%
\bibitem{Buchbinder:2014nia}
E.~Buchbinder and A.~Tseytlin,
\textit{``{The 1/N correction in the D3-brane description of circular Wilson
  loop at strong coupling}''},
\textsf{\doiref{10.1103/PhysRevD.89.126008}{Phys.Rev.~D89,~126008~(2014)}},
\texttt{\arxivref{1404.4952}}.
%
\bibitem{Buchel:2000cn}
A.~Buchel, A.~W.~Peet and J.~Polchinski,
\textit{``{Gauge dual and noncommutative extension of an N=2 supergravity
  solution}''},
\textsf{\doiref{10.1103/PhysRevD.63.044009}{Phys.Rev.~D63,~044009~(2001)}},
\texttt{\arxivref{hep-th/0008076}}.
%
\bibitem{Evans:2000ct}
N.~J.~Evans, C.~V.~Johnson and M.~Petrini,
\textit{``{The Enhancon and N=2 gauge theory: Gravity RG flows}''},
\textsf{JHEP~0010,~022~(2000)},
\texttt{\arxivref{hep-th/0008081}}.
%
\bibitem{Erickson:2000af}
J.~K.~Erickson, G.~W.~Semenoff and K.~Zarembo,
\textit{``{Wilson loops in N = 4 supersymmetric Yang-Mills theory}''},
\textsf{\doiref{10.1016/S0550-3213(00)00300-X}{Nucl.~Phys.~B582,~155~(2000)}},
\texttt{\arxivref{hep-th/0003055}}.
%
\bibitem{Drukker:2000rr}
N.~Drukker and D.~J.~Gross,
\textit{``{An exact prediction of N = 4 SUSYM theory for string theory}''},
\textsf{\doiref{10.1063/1.1372177}{J.~Math.~Phys.~42,~2896~(2001)}},
\texttt{\arxivref{hep-th/0010274}}.
%
\bibitem{Pilch:2003jg}
K.~Pilch and N.~P.~Warner,
\textit{``{Generalizing the N=2 supersymmetric RG flow solution of IIB
  supergravity}''},
\textsf{\doiref{10.1016/j.nuclphysb.2003.09.052}{Nucl.Phys.~B675,~99~(2003)}},
\texttt{\arxivref{hep-th/0306098}}.
%
\bibitem{Skenderis:2002vf}
K.~Skenderis and M.~Taylor,
\textit{``{Branes in AdS and p p wave space-times}''},
\textsf{\doiref{10.1088/1126-6708/2002/06/025}{JHEP~0206,~025~(2002)}},
\texttt{\arxivref{hep-th/0204054}}.
%

\bibitem{Schwarz:1983qr} 
J.~H.~Schwarz,
\textit{``{Covariant Field Equations of Chiral N=2 D=10 Supergravity}''},
\textsf{\doiref{doi:10.1016/0550-3213(83)90192-X}{Nucl.\ Phys.\ B {\bf 226}, 269 (1983).}}, 
  

\end{thebibliography}

\end{document}